\documentclass[fleqn,usenatbib]{mnras}

\usepackage[T1]{fontenc}
\DeclareRobustCommand{\VAN}[3]{#2}
\let\VANthebibliography\thebibliography
\def\thebibliography{\DeclareRobustCommand{\VAN}[3]{##3}\VANthebibliography}

\usepackage{graphicx}
\usepackage{amsmath}
\usepackage{amssymb}
\usepackage{txfonts}
\newcommand{\alpco}{\ensuremath{\alpha_{\rm CO}}}	
\newcommand{\alpcomw}{\ensuremath{\alpha_{\rm CO,MW}}}

\newcommand{\hi}{H\,{\sc i}}
\newcommand{\alpunits}{\ensuremath{M_{\odot}\,\mathrm{(K\,km\,s^{-1}\,pc^2)}^{-1}}}
\newcommand{\lunits}{\ensuremath{\mathrm{K\,km\,s^{-1}\,pc^2}}}

\title[CO observations of strong H$_2$-absorbing GRB hosts at $z>2$]{GRB host galaxies with strong H$_2$ absorption: CO-dark molecular gas at the peak of cosmic star formation}

\author[K. E. Heintz et al.]{K.~E.~Heintz,$^{1,2,3}$\thanks{E-mail: keh14@hi.is}
G.~Björnsson,$^{1}$
M.~Neeleman,$^{4}$
L.~Christensen,$^{2,3}$
J.~P.~U.~Fynbo,$^{2,3}$
P.~Jakobsson,$^{1}$
\newauthor
J.-K.~Krogager,$^{5}$
T.~Laskar,$^{6}$
C.~Ledoux,$^{7}$
G.~Magdis,$^{2,3}$
P.~M\o ller,$^{3,8}$ 
P.~Noterdaeme,$^{5}$
P.~Schady,$^{6}$
\newauthor
A.~de Ugarte Postigo,$^{9}$
F.~Valentino,$^{2,3}$ \& 
D.~Watson,$^{2,3}$
\\
$^{1}$Centre for Astrophysics and Cosmology, Science Institute, University of Iceland, Dunhagi 5, 107 Reykjav\'ik, Iceland \\
$^{2}$Cosmic Dawn Center (DAWN), Denmark,\\
$^{3}$Niels Bohr Institute, University of Copenhagen, Jagtvej 128, 2200 Copenhagen N, Denmark\\
$^{4}$Max-Planck-Institut f\"ur Astronomie, K\"onigstuhl 17, D-69117, Heidelberg, Germany\\
$^{5}$Institut d'Astrophysique de Paris, CNRS-SU, UMR\,7095, 98bis bd Arago, 75014 Paris, France \\
$^{6}$ Department of Physics, University of Bath, Claverton Down, Bath BA2 7AY, UK \\
$^{7}$European Southern Observatory, Alonso de C\'ordova 3107, Vitacura, Casilla 19001, Santiago, Chile\\
$^{8}$European Southern Observatory, Karl-Schwarzschildstrasse 2, D-85748 Garching bei M\"unchen, Germany\\
$^{9}$ Instituto de Astrofísica de Andalucía (IAA-CSIC), Glorieta de la Astronomía, s/n, E-18008, Granada, Spain
}

\date{Accepted XXX. Received YYY; in original form ZZZ}

\pubyear{2021}

\begin{document}
\label{firstpage}
\pagerange{\pageref{firstpage}--\pageref{lastpage}}
\maketitle

\begin{abstract}
We present a pilot search of 
CO emission in three 
H$_2$-absorbing, long-duration gamma-ray burst (GRB) host galaxies at $z\sim 2 - 3$.
We used the Atacama Large Millimeter/sub-millimeter Array (ALMA) to target the CO$(3-2)$ emission line and report non-detections for all three hosts. These are used to place limits on the host molecular gas masses, assuming a metallicity-dependent CO-to-H$_2$ conversion factor ($\alpha_{\rm CO}$). We find, $M_{\rm mol} < 3.5\times 10^{10}\,M_{\odot}$ (GRB\,080607), $M_{\rm mol} < 4.7\times 10^{11}\,M_{\odot}$ (GRB\,120815A), and $M_{\rm mol} < 8.9\times 10^{11}\,M_{\odot}$ (GRB\,181020A). The high limits on the molecular gas mass for the latter two cases are a consequence of their low stellar masses $M_\star$ ($M_\star \lesssim 10^{8}\,M_{\odot}$) and low gas-phase metallicities ($Z\sim 0.03\,Z_{\odot}$). The limit on the $M_{\rm mol}/M_\star$ ratio derived for GRB\,080607, however, is consistent with the average population of star-forming galaxies at similar redshifts and stellar masses.
We discuss the broader implications for a metallicity-dependent CO-to-H$_2$ conversion factor, and demonstrate that the canonical Galactic $\alpha_{\rm CO}$, will severely underestimate the actual molecular gas mass for all galaxies at $z>1$ with $M_\star < 10^{10}\,M_\odot$. To better quantify this we develop a simple approach to estimate the relevant $\alpha_{\rm CO}$ factor based only on the redshift and stellar mass of individual galaxies.
The elevated conversion factors will make these galaxies appear CO-``dark'' and difficult to detect in emission, as is the case for the majority of GRB hosts.
GRB spectroscopy thus offers a complementary approach to identify low-metallicity, star-forming galaxies with abundant molecular gas reservoirs at high redshifts that are otherwise missed by current ALMA surveys.
\end{abstract}

\begin{keywords}
gamma-ray burst: general -- galaxies: high-redshift, ISM, star formation -- ISM: molecules
\end{keywords}



\section{Introduction} \label{sec:intro}

Since long-duration gamma-ray bursts (GRBs) are linked to the death of massive stars \citep{Hjorth03,Stanek03,Woosley06,Cano17}, they are expected to trace star formation through cosmic time \citep{Wijers98,Kistler09,Robertson12,Greiner15}. GRB-selected galaxies therefore probe the underlying population of star-forming galaxies that are not biased towards the most luminous and massive galaxies unlike traditional emission-selected galaxy surveys. Moreover, the short-lived optical afterglows following GRBs are so bright that the plethora of absorption features that are imprinted from the interstellar medium (ISM) of the GRB host on the afterglow spectrum can be studied in detail \citep[e.g.,][]{Jakobsson04,Prochaska07,Vreeswijk07,Fynbo09}.  

\begin{table*}
	\centering
	\begin{minipage}{0.9\textwidth}
		\centering
		\caption{Sample properties of the H$_2$-bearing GRB host galaxies.}
		\begin{tabular}{lcccccccc}
		\noalign{\smallskip} \hline \hline \noalign{\smallskip}
		GRB & $z_{\rm GRB}$ & $\log N$(H$_2$) & $\log f_{\rm H_2}$ & [X/H] & $A_V$ & $\log M_\star$ & $L^\prime_{\rm CO(3-2)}$ & $\log M_{\rm mol}$   \\
		& & (cm$^{-2}$) & & & (mag) & ($M_\odot$) & ($10^9$\,K\,km\,s$^{-1}$\,pc$^2$) & ($M_\odot$) \\
		\noalign {\smallskip} \hline \noalign{\smallskip}
		080607 & 3.0363 & $21.20\pm 0.20$ & $-1.23\pm 0.24$ & $>-0.2$ & $2.58\pm 0.45$ & $10.45\pm 0.10$ & $<4.56$ & $<10.54$\\ 
		120815A & 2.3582 & $20.42\pm 0.08$ & $-1.39\pm 0.09$ & $-1.45\pm 0.03$ & $0.19\pm 0.04$ & $7.90\pm 0.40$ & $<0.55$ & $<11.67$\\ 
		181020A & 2.9379 & $20.40\pm 0.04$ & $-1.51\pm 0.06$& $-1.57\pm 0.06$ & $0.27\pm 0.02$ & $7.80\pm 0.40$ & $<0.71$ & $<11.95$ \\
			\noalign{\smallskip} \hline \noalign{\smallskip}
		\end{tabular}
		\label{tab:hosts}
	\end{minipage}
\end{table*}

After the first few afterglow spectra were obtained it was clear that GRB-host absorption systems typically probe sightlines with the highest \hi\ column densities of the so-called damped Lyman-$\alpha$ absorbers \citep[DLAs;][]{Vreeswijk04,Jakobsson06,Fynbo09}, related to their small impact parameters. DLAs provide the most effective and detailed probe of neutral gas in high-redshift galaxies, and contain most of the neutral gas at high redshift \citep{Noterdaeme09}. Given their direct link to star formation and the very high column densities of gas typically detected in GRB afterglow spectra, the low detection rate of molecular hydrogen H$_2$ (from the UV Lyman-Werner bands) was initially a puzzle \citep[e.g.,][]{Tumlinson07,Ledoux09}. The first detection of H$_2$ in a GRB absorber was observed in the remarkable afterglow spectrum of GRB\,080607 \citep{Prochaska09}\footnote{Though see also the tentative detection of H$_2$ in the afterglow of GRB\,060206 reported by \citet{Fynbo06}.}. Since then, eight more H$_2$-bearing GRB absorbers have been detected \citep{Kruhler13,DElia14,Friis15,Bolmer19,Heintz19}, largely owing to the extensive VLT/X-shooter GRB afterglow legacy survey \citep[XS-GRB;][]{Selsing19}. This GRB-selected sample of star-forming galaxies provides a unique way to study the molecular gas properties of high-$z$ galaxies in absorption.

To fully exploit the detailed information of intervening or host galaxy DLAs, it is important to study the association with their galaxy counterparts in emission. This has been done extensively for GRB hosts at UV to optical wavelengths \citep[e.g.,][]{Kruhler15,Arabsalmani18a,Corre18}. Similarly, surveys targeting GRB hosts at sub-mm wavelengths, in particular the molecular emission from carbon monoxide (CO) have advanced over the last few years \citep{Kohno05,Endo07,Hatsukade07,Hatsukade11,Hatsukade14,Hatsukade19,Hatsukade20,Stanway11,Stanway15,Arabsalmani18b,Michalowski16,Michalowski18,deUgartePostigo20}, largely due to the commissioning of the Atacama Large Millimeter/submillimeter Array (ALMA). Until now, however, only blind surveys or individual detections of CO emission in GRB host galaxies have been carried out.

In the pilot study presented here, we target a uniformly selected sample of GRB hosts, all at $z>2$, and identified solely on the basis of H$_2$ in absorption.
Contrary to the majority of absorption systems in quasar sightlines \citep[but see][]{Ranjan20}, we expect the bulk of the absorbing material to probe the ISM within the host galaxy, close to luminous regions of star-formation \citep[e.g.,][]{Fruchter06, Svensson10}.
These systems will thus allow us to study the molecular gas-phase in the central-most regions of star-forming galaxies, for which information can be obtained from the two complementary methods relying on molecular absorption and emission features. 

The present paper is structured as follows. In Sect.~\ref{sec:obs}, we present the sample criteria of our pilot study and the observational setup. Sect.~\ref{sec:res} presents the results in terms of the inferred molecular gas masses and how they compare to the overall population of GRB host galaxies. In Sect.~\ref{sec:disc}, we place the GRB hosts in the context of the underlying population of star-forming galaxies and discuss how they allow us to probe the elusive high-$z$, low-metallicity regime. In Sect.~\ref{sec:conc}, we summarize and conclude on our work.

\section{Sample and observations} \label{sec:obs}

\subsection{GRB hosts with strong H$_2$ absorption}

We targeted the three GRB-hosts with strongest H$_2$ absorption known to date, all at $z>2$. These are: GRB\,080607 \citep{Prochaska09}, GRB\,120815A \citep{Kruhler13}, and GRB\,181020A \citep{Heintz19}. All show H$_2$ column densities above $N{\rm (H_2)} > 10^{20}$\,cm$^{-2}$ and also constitute the GRB absorbers with the largest molecular-hydrogen fractions,
$f_{\rm H_2}$ \citep{Bolmer19,Heintz19}. 
GRB\,080607 also shows a high absorption-derived gas-phase metallicity consistent with Solar ([X/H] $> -0.2$) and significant dust extinction $A_V \sim 3$\,mag. On the contrary, GRBs\,120815A and 181020A show relatively low metallicities of [X/H] $\approx -1.5$ and modest extinction in the line-of-sight ($A_V = 0.2 - 0.3$\,mag).

The host-galaxy emission counterpart of GRB\,080607 is well detected in several bands \citep{Chen10,Wang12}. Here, we adopt the stellar population properties derived by \citet{Corre18}, inferring $\log (M_\star/M_\odot) = 10.45\pm 0.10$. For the other two, more recently detected GRBs, no host-galaxy counterpart has been identified yet. For these, we instead rely on previous work connecting DLAs to their emission counterparts \citep{Neeleman13,Moller13,Christensen14}. Following \citet{Arabsalmani15}, we assign an impact parameter of 2.3\,kpc and compute the predicted stellar masses based on the prescription by \citet[][their eq.~3]{Christensen14}. This yield stellar masses of $\log (M_\star/M_\odot) = 7.9\pm 0.4$ (GRB\,120815A) and $\log (M_\star/M_\odot) = 7.8\pm 0.4$ (GRB\,181020A), where the uncertainties are dominated by the internal scatter in $\log M_\star$.


\subsection{ALMA observations}

We observed the fields surrounding GRBs\,080607, 120815A, and 181020A, targeting the CO$(3-2)$ emission line as part of a dedicated ALMA Cycle 7 programme (ALMA Programme ID: 2019.1.00407.S, PI: Heintz). At the redshifts of the GRB hosts ($z\sim 2-3$), this line falls within the ALMA band 3 receiver. For each of the GRBs, we tuned one of the 1.875\,GHz spectral windows to the redshifted CO$(3-2)$ emission with a correlator setup yielding 960 channels with a width of 1.95~MHz. The remaining three bands were used to detect $\approx\,$95\,GHz continuum emission in the fields. Observations were done in a compact configuration with maximum baselines ranging between 300 and 500\,m depending on the observing block. Total on-source integration times were 3.1, 2.4, and 1.6 hours for GRBs\,080607, 120815A, and 181020A, respectively. 

The raw data were calibrated using the ALMA Pipeline, which is part of the Common Astronomy Software Application package \citep[CASA;][]{McMullin07}. After the initial calibration, additional manual data editing was performed using the flagging routines within CASA. Both the continuum image as well as the spectral cube centered around the redshifted CO$(3-2)$ line were obtained using the task \textit{tclean} within CASA by applying natural weighting to maximize sensitivity to point sources. This resulted in spatial resolutions of $(4.0" \times 3.6")$, $(2.3" \times 2.1")$, and $(3.8" \times 3.5")$ for GRBs\,080607, 120815A, and 181020A, respectively.
For all GRBs, we also Hanning-smoothed the spectral cube to a velocity resolution of 25 and 100\,km\,s$^{-1}$. The resultant sensitivities per 100\,km\,s$^{-1}$ are 94, 85, and $94\,\mu$Jy\,beam$^{-1}$ for the spectral cube and 5.8, 5.7, and $6.3\,\mu$Jy\,beam$^{-1}$ at 92.6, 96.0, and 92.9\,GHz for the continuum image, for GRBs\,080607, 120815A, and 181020A, respectively. 

\begin{figure}
\begin{center}
\includegraphics[width=8.5cm]{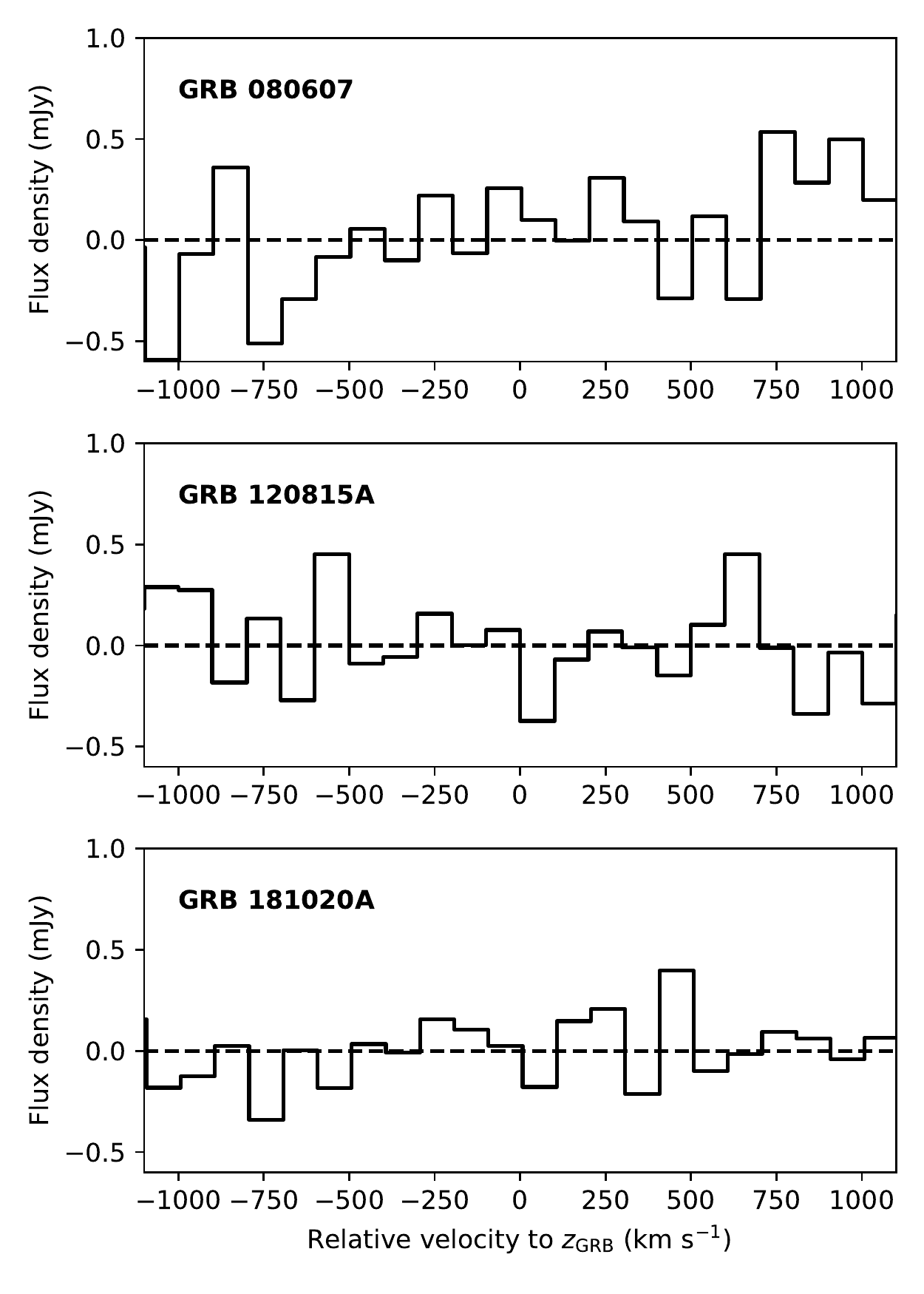}
\caption{CO flux density as a function of velocity, where $v_{\rm rel} = 0$\,km\,s$^{-1}$ corresponds to the redshift of the absorbers, $z_{\rm GRB}$. None of the spectra show detection of CO$(3-2)$ lines at the redshifts of the GRBs.}
\label{fig:almaspec}
\end{center}
\end{figure}



\section{Results} \label{sec:res}

We searched both the 25 and 100\,km\,s$^{-1}$ channel width ALMA spectral data cubes for emission originating from CO($3-2$) at the relevant redshifts for the GRBs in our sample. No line emission is detected at the position of any of the GRBs
(but see, e.g., Neeleman et al., in prep., for a detection of CO$(2-1)$ from the strong intervening Mg\,{\sc ii} absorber towards GRB\,120815A). We also did not detect continuum emission at the positions of any of the three GRBs, but there appears to be continuum emission from an unrelated galaxy in the field of GRB\,120815A. 


We extracted 1D spectra from the ALMA spectral data cube centred on the positions of the GRB afterglows or host galaxies, shown in Fig.~\ref{fig:almaspec}. We derived $3\sigma$ upper limits on the velocity-integrated flux densities of $<0.09$\,Jy\,km\,s$^{-1}$ (GRB\,080607), $<0.018$\,Jy\,km\,s$^{-1}$ (GRB\,120815A), and $<0.016$\,Jy\,km\,s$^{-1}$ (GRB\,181020A), assuming line widths for the emission-line profiles of ${\rm FWHM} = 300\,{\rm km\,s^{-1}}$ (GRB\,080607) and ${\rm FWHM} = 50\,{\rm km\,s^{-1}}$ (GRBs\,120815A and 181020A), appropriate for galaxies in their given mass ranges \citep[e.g.,][]{Tiley16}. 
We then derived the corresponding CO$(3-2)$ line luminosities \citep[following Eq. 3 from][]{Solomon05} of $L^\prime_{\rm CO(3-2)} < 4.56\times 10^{9}$\,K\,km\,s$^{-1}$\,pc$^2$ (GRB\,080607), $L^\prime_{\rm CO(3-2)} < 5.51\times 10^{8}$\,K\,km\,s$^{-1}$\,pc$^2$ (GRB\,120815A), and $L^\prime_{\rm CO(3-2)} < 7.12\times 10^{8}$\,K\,km\,s$^{-1}$\,pc$^2$ (GRB\,181020A).

We converted the measured CO$(3-2)$ line luminosities into total molecular gas masses assuming a line ratio of $r_{31} = L^\prime_{\rm CO(3-2)} / L^\prime_{\rm CO(1-0)} = 0.57$ \citep[which is the observed average for $z>1$ star-forming galaxies;][]{DessaugesZavadsky15} 
and adopting a metallicity-dependent CO-to-H$_2$ conversion factor:

\begin{equation}
    \alpha_{\rm CO}(Z) = 4.5 \times (Z/Z_{\odot})^{-1.40}~M_\odot\,({\rm K\,km\,s}^{-1}\,{\rm pc}^2)^{-1},
\label{eq:aCO}
\end{equation}
following \citet{HeintzWatson20}. This \alpco-metallicity relation is the average between the locally-derived \citep{Israel97,Leroy11,Bolatto13,Amorin16} and high-redshift \citep{Genzel12} inferred relations, and is calibrated to galaxies at $z>1$. This yields upper limits on the molecular gas masses of $\log (M_{\rm mol}/M_{\odot}) < 10.54, 11.67,$ and 11.95, for the hosts of GRBs\,080607, 120815A, and 181020A, respectively.
All the values derived in this section are summarized in Table~\ref{tab:hosts}.

\begin{figure}
\begin{center}
\includegraphics[width=8.5cm]{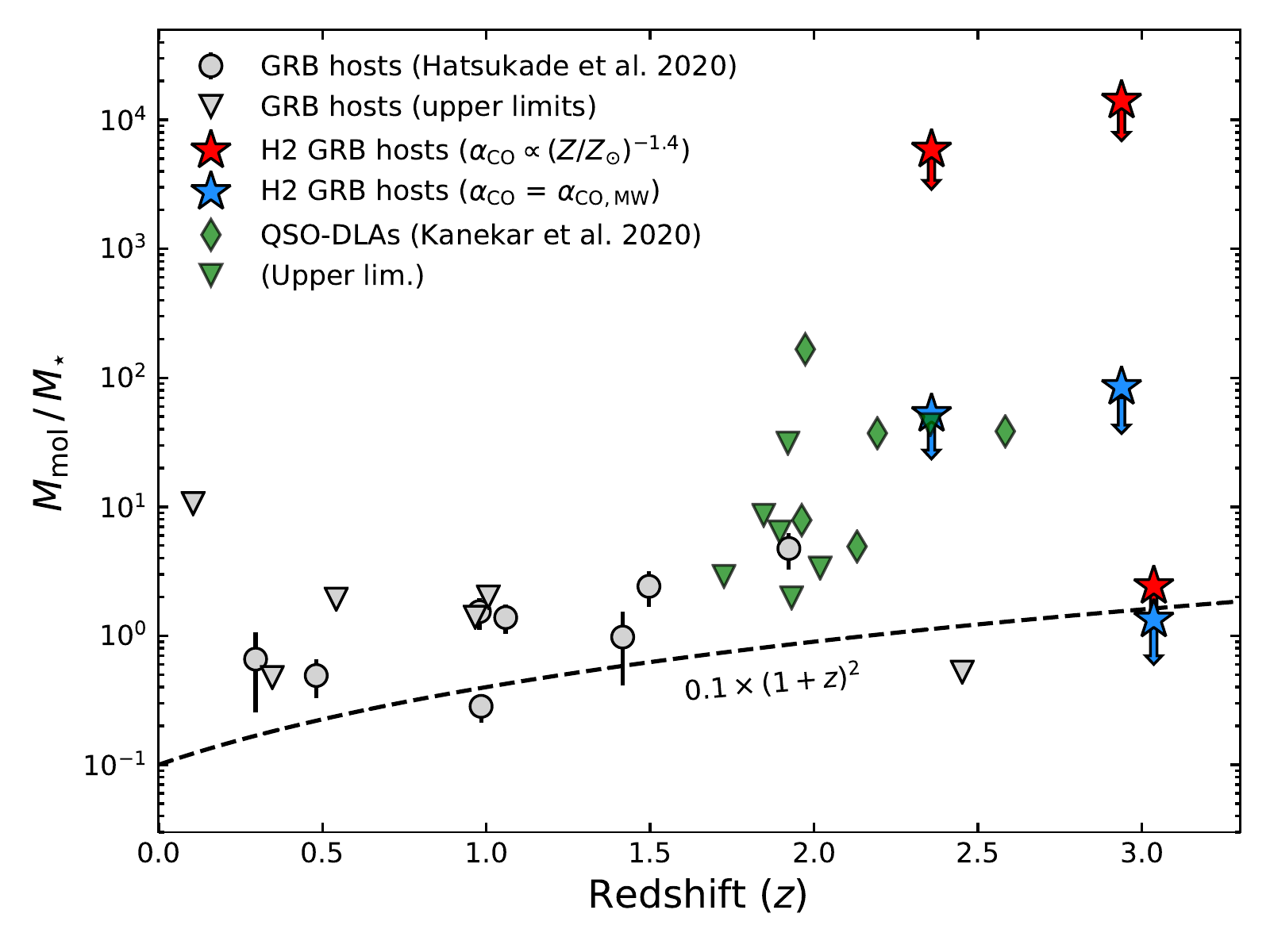}
\caption{Molecular gas-to-stellar mass ratio for our three GRB host galaxy sample (star-symbols). Shown are the ratios assuming a MW-like CO-to-H$_2$ conversion factor (blue) and using the \alpco-metallicity relation (eq.~\ref{eq:aCO}, red). Detections and upper limits for the GRB host sample from \citet{Hatsukade20} and quasar DLAs \citep{Kanekar20} are also shown. The dashed curve shows the evolutionary track broadly characterizing $M_\star > 10^{10}\,M_\odot$ main-sequence star-forming galaxies \citep{Geach11}.}
\label{fig:mmol}
\end{center}
\end{figure}

In Fig.~\ref{fig:mmol} we show the upper limits of the molecular gas-to-stellar mass ratio of the three GRB hosts. In each case, the molecular gas mass ratio is determined by using either the above \alpco-metallicity relation and the host absorption-derived metallicities, or by assuming a constant MW-like conversion factor of $\alpcomw = 4.3$\,\alpunits\ \citep{Bolatto13}. For comparison, we overplot the recent compilation of CO observations of GRB host galaxies from \citet{Hatsukade20}, spanning the redshift range $z=0.0-2.5$. In addition, we show the track that broadly characterizes the molecular gas mass evolution of $M_\star > 10^{10}M_\odot$ main-sequence star-forming galaxies, $M_{\rm mol}/M_\star = 0.1\times(1+z)^2$ \citep{Geach11,Carilli13}. 

While the ratio of molecular gas mass to stellar mass of the hosts of GRBs\,120815A and 181020A are poorly constrained, the host galaxy of GRB\,080607 is only marginally consistent with that expected for typical star-forming galaxies at similar redshifts and stellar masses. 
This is in stark contrast with the population of DLAs at $z\gtrsim 2$ observed in quasar sightlines, shown as the green symbols in Fig.~\ref{fig:mmol}, which overall show a significant excess of molecular gas \citep{Neeleman18,Kanekar20}.
Assuming a lower CO excitation of $r_{31} = 0.4$ \citep[mostly representative of the $z\lesssim 2$ population;][]{Boogaard20}, however, results in a consistent limit on the molecular-to-stellar mass content. The high upper limits on the $M_{\rm mol}/M_\star$ fractions of the hosts of GRBs\,120815A and 181020A are mainly due to their low metallicities requiring high CO-to-H$_2$ conversion factors to infer their molecular gas content. These host galaxies, however, also have lower stellar masses than the typical star-forming galaxies probed in CO at similar redshifts \citep[e.g.,][]{Tacconi13,Tacconi18}. In the next section, we will explore these particular high-redshift, low-mass, and low-metallicity hosts in context to the underlying field-galaxy population.



\section{Discussion} \label{sec:disc}

\subsection{Field environment of GRB hosts}

Studies of absorption-selected galaxies in the line-of-sight toward bright background quasars have revealed that many of these DLAs are found in environments with other nearby galaxies, both at low- \citep{Kacprzak10,Rahmani18} and high-redshifts \citep{Moller93,Francis93,Fynbo03}. These DLA counterparts have mostly been detected based on strong rest-frame optical emission lines, or from Lyman-$\alpha$ emission from the galaxy counterparts, but are now also being increasingly detected in CO \citep{Klitsch18,Fynbo18}. With the data presented here we can examine in an independent way whether GRB-DLAs also appear to be part of larger galaxy complexes. 

Each ALMA cube covers $\approx 45''$ subtended on the sky, which corresponds to 350\,kpc and 375\,kpc at the target redshifts ($z=3$ and $z=2.3$), respectively. We do not detect emission from CO$(3-2)$ in galaxies within the $\approx \pm 1500$\,km/s covered by the data cube, corresponding to $z=\pm 0.02$. We thus do not find evidence for galaxy clustering in this (albeit small) sample of CO-surveyed GRB hosts. The quasar-DLA bias towards CO-emitting galaxy groups could partly be explained by the preferential high-metallicity these systems were selected on. Consequently, absorption-line features in quasar sightlines could therefore be influenced by these more populated galaxy environments \citep[e.g.,][]{Hamanowicz20}, rather than tracing the line-of-sight through a single galactic disk \citep{Fynbo18}. GRB sightlines could therefore provide a cleaner probe of absorption-derived quantities and correlations \citep[e.g.,][]{Arabsalmani15,Arabsalmani18a}. 


\subsection{Mass and redshift dependence of $\alpco$}

One of the most promising aspects of studying the CO emission associated to GRB hosts, is that it allows us to probe the molecular gas content in low-metallicity galaxies which are otherwise missed by field-selected surveys.
As metallicity decreases both with increasing redshift and decreasing galaxy mass, we expect \alpco\ to show a strong mass and redshift dependence. In fact, even a metallicity of $Z/Z_{\odot} \approx 10\%$ would imply an \alpco\ value more than an order of magnitude higher than the average Milky Way conversion factor \alpcomw. The reason why this is still largely applied to high-redshift galaxies, is due to the difficulty in measuring the gas-phase metallicity in low-metallicity, low-luminosity galaxies. 
Below, we aim to improve on this and present a simple relation that conveniently expresses the CO-to-H$_2$ conversion factor, in addition to the $M_{\rm mol}/M_\star$ ratio, as a function of stellar mass and redshift. 

Starting from the metallicity-dependent \alpco\ relation (Eq.~\ref{eq:aCO}), we can connect \alpco\ directly to the galaxy stellar mass at any given redshift via mass-metallicity relations. For the latter, \cite{Savaglio05} provided a convenient fit, valid for a wide mass and redshift ranges, that has been improved upon by \cite{Genzel15} who combined metallicity prescriptions from four different studies. The latter paper, however, introduces a solar-metallicity cut-off at low redshift, that is too limiting for our purposes. Instead, we start with the results of \cite{Maiolino08}, one of the four prescriptions used by \cite{Genzel15}, that provide at each redshift bin a mass-metallicity relation:
\begin{equation}
    12+\log({\rm O/H}) = -0.0864(\log M_{\star} - \log M_z)^2 + K_0.
\label{eq:oh}
\end{equation}
Here, $M_{\star}$ is the stellar mass and $\log M_z$ and $K_0$ are constants in each bin, their values given in \citet[Table 5]{Maiolino08}. We fit a function of $(1+z)$ to these factors and find that $\log M_z = 2.59\times \log(1+z) + 11.05$ and $K_0 = 8.9$, provides a good representation of their redshift evolution. The $K_0$-values can also be well fitted by a $2^{\rm nd}$-order polynomial in $z$, but this significantly under-predicts the metallicity at $z>3$ compared to other calibrators. With the above form for $\log M_z$ and a constant $K_0$, we allow for super-solar metallicities and find metallicity-values that are bracketed by 
the mass-metallicity calibrations of \cite{Savaglio05} and \cite{Genzel15}, at all but the lowest masses and lowest redshifts. We include an uncertainty of 0.2 dex in this mass-metallicity relation, representing the typical dispersion around this relation. 

\begin{figure}
\begin{center}
\includegraphics[width=\columnwidth]{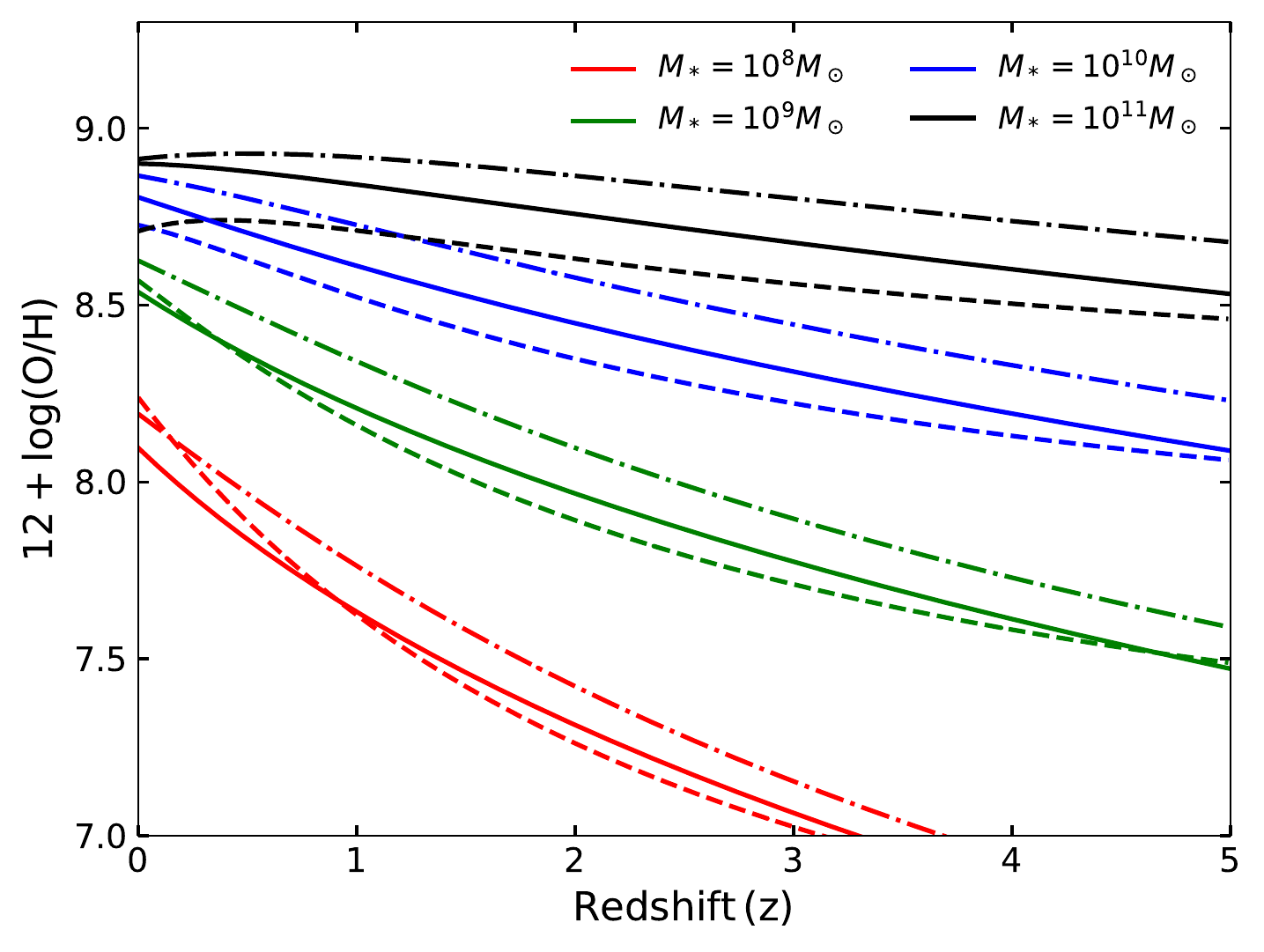}
\caption{Metallicity as a function of redshift for four different galaxy stellar masses. Our fit to the relations in \citet{Maiolino08} is shown with solid curves, while \citet{Savaglio05} is dashed-dotted and \citet{Genzel15} is dashed. Note that our fit lies in between the other two at all redshifts, except for the lowest masses at the lowest redshifts.}
\label{fig:massz}
\end{center}
\end{figure}

\begin{figure*}
\begin{center}
\includegraphics[width=\columnwidth]{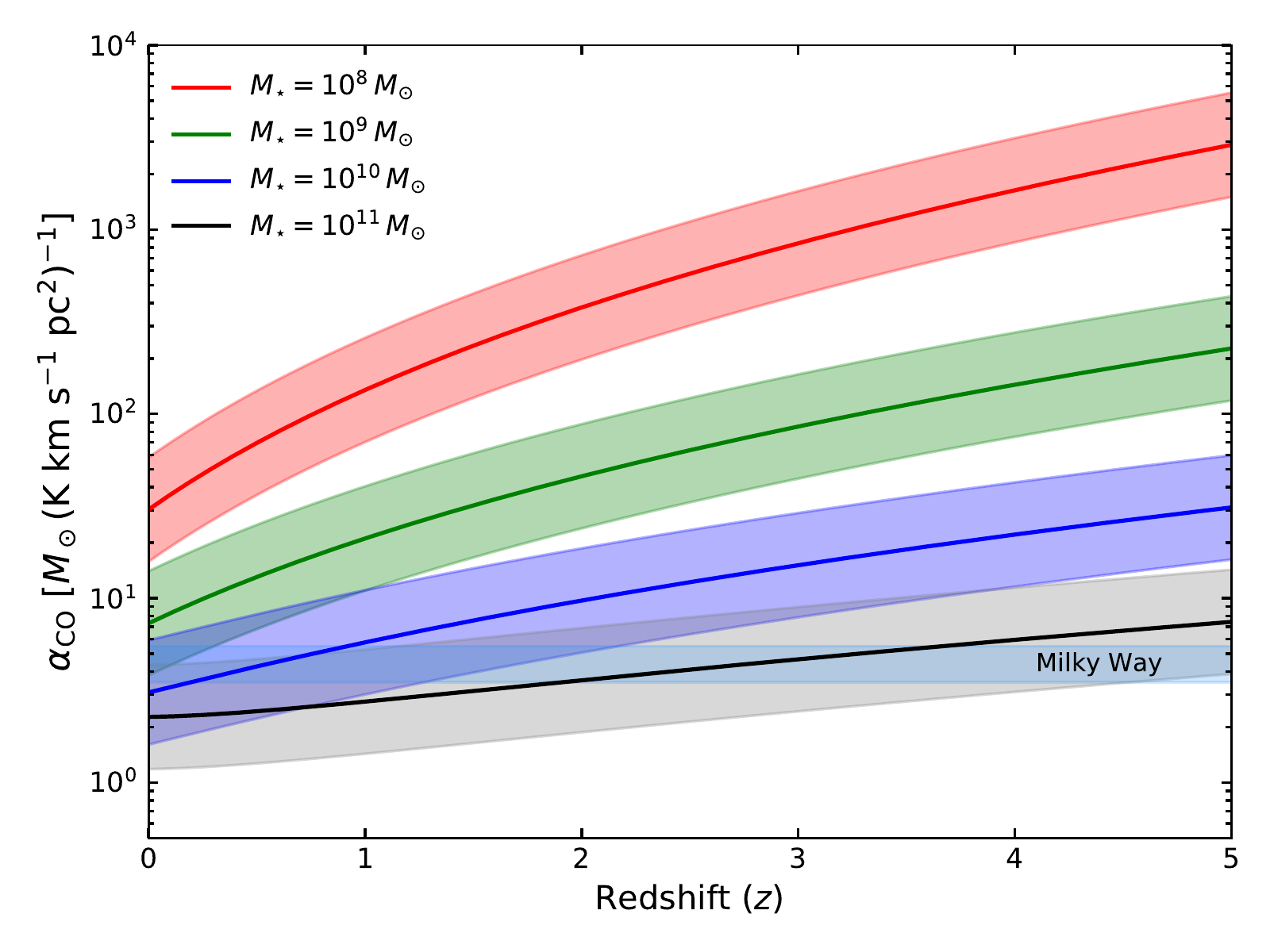}
\includegraphics[width=\columnwidth]{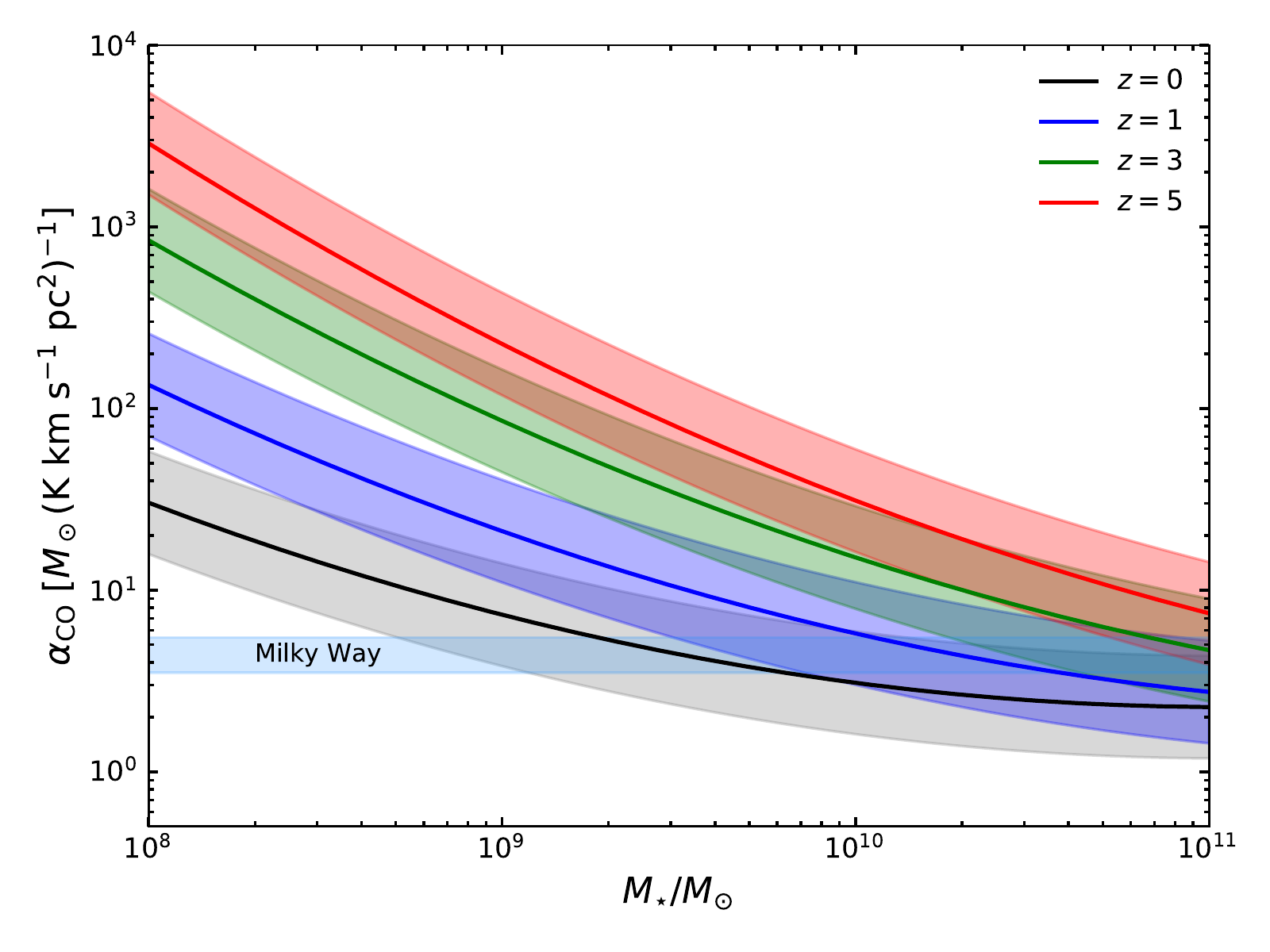}
\caption{The evolution of the molecular gas mass to CO($1-0$) luminosity ratio, \alpco, for regular star-forming galaxies typical galaxy scaling relations. The left panel shows the evolution as a function of redshift for selected galaxy stellar masses and the right panel shows the corresponding evolution as a function of stellar mass for the given redshifts. 
The shaded error region on each curve represents the combined errors from the mass-metallicity relation and \alpco-metallicity relation (see main text for more details). The average value observed in the Milky Way is shown with the blue horizontal band in both panels.}
\label{fig:acoevol}
\end{center}
\end{figure*}

Connecting the \alpco-metallicity relation with the mass-metallicity calibration derived, we obtain a simple expression for \alpco:
\begin{equation}
    \log \alpha_{\rm CO} = 0.121\left(\log M_{\star} - \log M_{z}\right)^2 + 0.359,
\end{equation}
depending only on the stellar mass, $M_{\star}$, of a galaxy at redshift $z$. Again, $\log M_z = 2.59\times \log(1+z) + 11.05$, representing the characteristic turn-over stellar mass at a given redshift. These relations allow us to estimate directly the appropriate molecular gas mass conversion factors for main-sequence star-forming galaxies at any given redshift and stellar mass for which the mass-metallicity calibration is accurate. 

Based on this combined relation, we explore how \alpco\ evolves in typical star-forming galaxies as a function of redshift and stellar mass. We show these evolutionary tracks in 
Fig.~\ref{fig:acoevol} for a set of representative redshifts and stellar masses, namely $z=0,1,3,5$ and $M_{\star} = 10^{8},10^{9},10^{10},10^{11}\,M_{\odot}$.
The shaded error regions on each curve represent the combined errors from the typical dispersion around the mass-metallicity relation ($\sigma_{\rm MZ} = 0.2$ dex) and the scatter in the best-fit \alpco-metallicity relation. 
In the figure, we also compare the evolution of \alpco\
to the average values observed in the Milky Way and in local galaxies (i.e.\ equivalent to solar metallicities) of $\alpcomw = 3.5 - 5.5$\,\alpunits\ \citep{Bolatto13} 

As suggested by the mass-metallicity relation (eq.~\ref{eq:oh}), the evolution of \alpco\ depends strongly on both galaxy mass and redshift. We find that the Galactic conversion factor \alpcomw\ is a suitable approximation for massive galaxies, with $M_\star \sim 10^{11}\,M_{\odot}$, at all redshifts in the range $z = 0 - 5$. Similarly, galaxies at $z\sim 0$ with stellar masses in the range $M_{\star} = 10^{9} - 10^{11}\,M_{\odot}$ also show conversion factors consistent with that observed in the Milky Way within the uncertainties. However, we do recover an overall increase in \alpco\ at decreasing stellar masses for fixed redshifts as expected. Galaxies with $M_{\star} < 10^{9}\,M_{\odot}$ and also most galaxies at $z > 1$ (except the most massive ones, $M_{\star} \sim 10^{11}\,M_{\odot}$), i.e. the majority of GRB hosts, show \alpco\ values significantly higher than \alpcomw. For instance, a galaxy with $M_\star = 10^{8}\,M_{\odot}$ at $z \sim 0$ will have a conversion factor of $\alpco = 30.5$\,\alpunits,
almost an order of magnitude higher than the average value observed in the Milky Way. At higher redshifts, where CO is now being increasingly detected \citep[e.g.,][]{Tacconi10,Tacconi13,Tacconi18,Walter11,Decarli16,Bothwell17,Aravena19,Pavesi18,Riechers19,Valentino18,Valentino20}, even massive galaxies with $M_{\star} \sim 10^{10}\,M_{\odot}$ at $z = 2.5$ will have conversion factors of $\alpco = 12.3$\,\alpunits,
exceeding the average Galactic value by a factor of $2-3$.

The redshift- and mass-dependent evolution of \alpco\ can now be used to shed light on the observed $M_{\rm mol}/M_\star$-redshift relation. As discussed in the previous section, it has been shown that this ratio approximately follows $M_{\rm mol}/M_\star \propto (1+z)^{2.5}$, for star-forming galaxies \citep{Tacconi18}. In the following analysis we assume a constant value of $L^\prime_{\rm CO(1-0)} = 2\times 10^{10}$\,\lunits, which is the average luminosity of galaxies at $z\sim 1-3$ derived from the PHIBSS sample \citep[corrected by $r_{31}$;][]{Tacconi13,Tacconi18}. We caution that the CO luminosity has been found to be strongly correlated with the stellar mass of each galaxy \citep[e.g.][]{Inami20}, and that this analysis should only be treated as a simple model to explore the evolution of $M_{\rm mol}/M_\star$ with redshift. 

\begin{figure}
\begin{center}
\includegraphics[width=8.9cm]{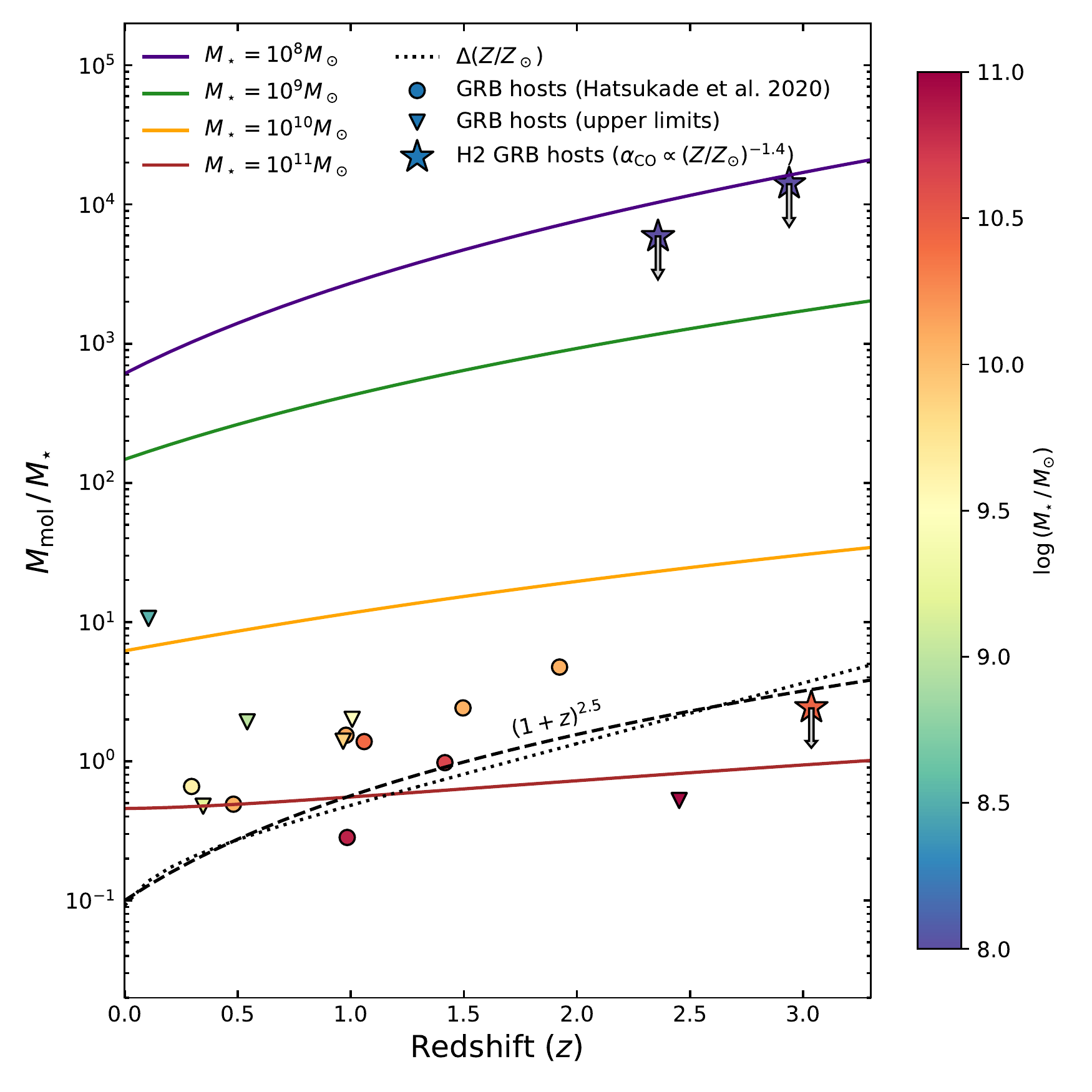}
\caption{Same as Fig.~\ref{fig:acoevol}, but with curves of constant stellar mass overplotted. Only the upper-limits using the \alpco-metallicity relation (Eq.~\ref{eq:aCO}), are shown for the three GRB hosts from this work. All plotted data points are color-coded according to stellar mass. In addition, the dotted curve shows the expected evolution for a 0.2 dex metallicity decrease per unit redshift, starting at $\approx 1.5\times Z_{\odot}$.}
\label{fig:mmolmass}
\end{center}
\end{figure}

In Fig.~\ref{fig:mmolmass}, we plot the resulting $M_{\rm mol}/M_\star$-curves for a range of stellar masses, $M_{\star} = 10^{8}, 10^{9}, 10^{10}, 10^{11}\,M_{\odot}$, covering redshifts from $z=0-3$, with the GRB hosts overplotted for reference. For the hosts of GRBs\,120815A and 181020A, the upper limits are weaker than expected from their host stellar masses, whereas the limit on the molecular gas mass in GRB\,080607 is close to that expected from its host stellar mass. Overall, most of the CO detections in GRB hosts do fall between the curves with $M_{\star} = 10^{10} - 10^{11}\,M_{\odot}$ \citep{Hatsukade20}, as expected from their stellar masses, though some hosts appear to show significant molecular deficits (in particular at $z \lesssim 1$).

Finally, in Fig.~\ref{fig:mmolmass}, we show the expected $M_{\rm mol}/M_\star$ evolution track assuming a 0.2 dex metallicity decrease per unit redshift \citep[as observed for absorption-selected galaxies, e.g.][]{Rafelski12,DeCia18}. The trend start at $Z= 1.5 Z_{\odot}$ at $z=0$, and we again assume $L^\prime_{\rm CO(1-0)} = 2\times 10^{10}$\,\lunits\ and the mass-metallicity and $\alpco$-mass relations described above (shown as the dotted line in the figure). It is clear that this evolution track approximately follows the relation, $M_{\rm mol}/M_\star \propto (1+z)^{2.5}$, that has been considered to broadly describe the evolution of the more massive galaxies, $M_\star > 10^{10}M_\odot$. While the stellar mass is often thought to be the main factor regulating the metallicity of galaxies, we here show that the metallicity evolution of typical massive star-forming galaxies might in fact be the main driver of the observed molecular gas fraction, $M_{\rm mol}/M_\star$, in galaxies through cosmic time. 


\section{Conclusions} \label{sec:conc}

We here presented a pilot survey targeting the CO emission counterparts of the host galaxies of strong H$_2$-absorbing GRBs at $z\sim 2-3$. We did not detect the redshifted CO$(3-2)$ line in any of the three GRBs (GRBs\,080607, 120815A, and 181020A). Assuming typical line ratios observed for star-forming galaxies at similar redshifts and a metallicity-dependent CO-to-H$_2$ conversion factor, we derived upper limits on the molecular gas masses of $\log ( M_{\rm mol}/M_\star ) < 10.54, 12.09,$ and 12.47, for the hosts of GRBs\,080607, 120815A, and 181020A, respectively.

To place these H$_2$-selected GRB host galaxies into context, we compared them to the most recent compilation of GRB hosts with CO observations \citep[][and references therein]{Hatsukade20}. First, the systems presented here expand the redshift range for which CO observations of GRBs have been obtained. Then, we examined them in terms of their molecular gas and stellar mass contents. While no strong conclusion can be inferred for the molecular-gas content of the hosts of GRBs\,120815A and 181020A, we demonstrate that the host galaxy of GRB\,080607 is globally deficient in molecules or ``CO-dark''. This is surprising given the high metallicity ([X/H] $>-0.2$) and H$_2$ abundance ($N({\rm H}_2) = 10^{21.2}$\,cm$^{-2}$) inferred from absorption, and the relatively high stellar mass inferred from emission ($\log (M_\star/M_\odot) = 10.45$).  

Motivated by the high inferred limits on the molecular gas masses of the hosts of GRBs\,120815A and 181020A, resulting from their low absorption-derived metallicities and a metallicity-dependent CO-to-H$_2$ conversion factor \alpco, we derived evolutionary tracks for \alpco\ as a function of redshift and stellar mass. We found that while the Galactic conversion factor \alpcomw\ is a suitable for massive galaxies with $M_\star \sim 10^{11}\,M_{\odot}$ at $z\sim 0 - 5$, lower mass galaxies will show significantly higher \alpco\ values at all redshifts (by up to several orders of magnitudes). This will hamper the detection probability of CO in most star-forming galaxies at $z>1$, since even large molecular gas reservoirs will show limited CO emission at these redshifts. 

We demonstrated in the pilot survey presented here, resulting in non-detections of CO emission from GRB host galaxies, that the most feasible way to identify and study the molecular gas reservoirs in high-redshift, low-metallicity galaxies is through the detection of H$_2$ in absorption. Due to the metallicity-dependent CO-to-H$_2$ conversion factor \alpco, these galaxies that otherwise show strong H$_2$ absorption will be too faint to be detected in emission using typical molecular gas tracers such as CO.

\section*{Acknowledgements}

KEH and PJ acknowledge support by a Project Grant (162948--051) from The Icelandic Research Fund. MN acknowledges support from ERC Advanced grant 740246 (Cosmic{\verb|_|}Gas). JPUF thanks the Carlsberg Foundation for support. The Cosmic DAWN center is funded by the DNRF.
PN and JKK acknowledge support from the French {\sl Agence Nationale de la Recherche} under grant ANR 17-CE31-0011-01 / project ``HIH2'' (PI Noterdaeme).


\section*{Data availability statement}

The raw ALMA data is publically available through the ALMA science archive. The reduced spectral cubes and source codes for the figures and tables presented in this manuscript are available from the corresponding author upon reasonable request.\\



\bibliographystyle{mnras}
\bibliography{ref} 

\begin{thebibliography}{}
\makeatletter
\relax
\def\mn@urlcharsother{\let\do\@makeother \do\$\do\&\do\#\do\^\do\_\do\%\do\~}
\def\mn@doi{\begingroup\mn@urlcharsother \@ifnextchar [ {\mn@doi@}
  {\mn@doi@[]}}
\def\mn@doi@[#1]#2{\def\@tempa{#1}\ifx\@tempa\@empty \href
  {http://dx.doi.org/#2} {doi:#2}\else \href {http://dx.doi.org/#2} {#1}\fi
  \endgroup}
\def\mn@eprint#1#2{\mn@eprint@#1:#2::\@nil}
\def\mn@eprint@arXiv#1{\href {http://arxiv.org/abs/#1} {{\tt arXiv:#1}}}
\def\mn@eprint@dblp#1{\href {http://dblp.uni-trier.de/rec/bibtex/#1.xml}
  {dblp:#1}}
\def\mn@eprint@#1:#2:#3:#4\@nil{\def\@tempa {#1}\def\@tempb {#2}\def\@tempc
  {#3}\ifx \@tempc \@empty \let \@tempc \@tempb \let \@tempb \@tempa \fi \ifx
  \@tempb \@empty \def\@tempb {arXiv}\fi \@ifundefined
  {mn@eprint@\@tempb}{\@tempb:\@tempc}{\expandafter \expandafter \csname
  mn@eprint@\@tempb\endcsname \expandafter{\@tempc}}}

\bibitem[\protect\citeauthoryear{{Amor{\'\i}n}, {Mu{\~n}oz-Tu{\~n}{\'o}n},
  {Aguerri}  \& {Planesas}}{{Amor{\'\i}n} et~al.}{2016}]{Amorin16}
{Amor{\'\i}n} R.,  {Mu{\~n}oz-Tu{\~n}{\'o}n} C.,  {Aguerri} J.~A.~L.,
  {Planesas} P.,  2016, \mn@doi [\aap] {10.1051/0004-6361/201526397}, \href
  {https://ui.adsabs.harvard.edu/abs/2016A&A...588A..23A} {588, A23}

\bibitem[\protect\citeauthoryear{{Arabsalmani}, {M{\o}ller}, {Fynbo},
  {Christensen}, {Freudling}, {Savaglio}  \& {Zafar}}{{Arabsalmani}
  et~al.}{2015}]{Arabsalmani15}
{Arabsalmani} M.,  {M{\o}ller} P.,  {Fynbo} J. P.~U.,  {Christensen} L.,
  {Freudling} W.,  {Savaglio} S.,   {Zafar} T.,  2015, \mn@doi [\mnras]
  {10.1093/mnras/stu2138}, \href
  {https://ui.adsabs.harvard.edu/abs/2015MNRAS.446..990A} {446, 990}

\bibitem[\protect\citeauthoryear{{Arabsalmani} et~al.,}{{Arabsalmani}
  et~al.}{2018a}]{Arabsalmani18a}
{Arabsalmani} M.,  et~al., 2018a, \mn@doi [\mnras] {10.1093/mnras/stx2451},
  \href {https://ui.adsabs.harvard.edu/abs/2018MNRAS.473.3312A} {473, 3312}

\bibitem[\protect\citeauthoryear{{Arabsalmani} et~al.,}{{Arabsalmani}
  et~al.}{2018b}]{Arabsalmani18b}
{Arabsalmani} M.,  et~al., 2018b, \mn@doi [\mnras] {10.1093/mnras/sty194},
  \href {https://ui.adsabs.harvard.edu/abs/2018MNRAS.476.2332A} {476, 2332}

\bibitem[\protect\citeauthoryear{{Aravena} et~al.,}{{Aravena}
  et~al.}{2019}]{Aravena19}
{Aravena} M.,  et~al., 2019, \mn@doi [\apj] {10.3847/1538-4357/ab30df}, \href
  {https://ui.adsabs.harvard.edu/abs/2019ApJ...882..136A} {882, 136}

\bibitem[\protect\citeauthoryear{{Bolatto}, {Wolfire}  \& {Leroy}}{{Bolatto}
  et~al.}{2013}]{Bolatto13}
{Bolatto} A.~D.,  {Wolfire} M.,   {Leroy} A.~K.,  2013, \mn@doi [\araa]
  {10.1146/annurev-astro-082812-140944}, \href
  {https://ui.adsabs.harvard.edu/abs/2013ARA&A..51..207B} {51, 207}

\bibitem[\protect\citeauthoryear{{Bolmer} et~al.,}{{Bolmer}
  et~al.}{2019}]{Bolmer19}
{Bolmer} J.,  et~al., 2019, \mn@doi [\aap] {10.1051/0004-6361/201834422}, \href
  {https://ui.adsabs.harvard.edu/abs/2019A&A...623A..43B} {623, A43}

\bibitem[\protect\citeauthoryear{{Boogaard} et~al.,}{{Boogaard}
  et~al.}{2020}]{Boogaard20}
{Boogaard} L.~A.,  et~al., 2020, arXiv e-prints, \href
  {https://ui.adsabs.harvard.edu/abs/2020arXiv200904348B} {p. arXiv:2009.04348}

\bibitem[\protect\citeauthoryear{{Bothwell} et~al.,}{{Bothwell}
  et~al.}{2017}]{Bothwell17}
{Bothwell} M.~S.,  et~al., 2017, \mn@doi [\mnras] {10.1093/mnras/stw3270},
  \href {https://ui.adsabs.harvard.edu/abs/2017MNRAS.466.2825B} {466, 2825}

\bibitem[\protect\citeauthoryear{{Cano}, {Wang}, {Dai}  \& {Wu}}{{Cano}
  et~al.}{2017}]{Cano17}
{Cano} Z.,  {Wang} S.-Q.,  {Dai} Z.-G.,   {Wu} X.-F.,  2017, \mn@doi [Advances
  in Astronomy] {10.1155/2017/8929054}, \href
  {https://ui.adsabs.harvard.edu/abs/2017AdAst2017E...5C} {2017, 8929054}

\bibitem[\protect\citeauthoryear{{Carilli} \& {Walter}}{{Carilli} \&
  {Walter}}{2013}]{Carilli13}
{Carilli} C.~L.,  {Walter} F.,  2013, \mn@doi [\araa]
  {10.1146/annurev-astro-082812-140953}, \href
  {https://ui.adsabs.harvard.edu/abs/2013ARA&A..51..105C} {51, 105}

\bibitem[\protect\citeauthoryear{{Chen} et~al.,}{{Chen} et~al.}{2010}]{Chen10}
{Chen} H.-W.,  et~al., 2010, \mn@doi [\apjl] {10.1088/2041-8205/723/2/L218},
  \href {https://ui.adsabs.harvard.edu/abs/2010ApJ...723L.218C} {723, L218}

\bibitem[\protect\citeauthoryear{{Christensen}, {M{\o}ller}, {Fynbo}  \&
  {Zafar}}{{Christensen} et~al.}{2014}]{Christensen14}
{Christensen} L.,  {M{\o}ller} P.,  {Fynbo} J.~P.~U.,   {Zafar} T.,  2014,
  \mn@doi [\mnras] {10.1093/mnras/stu1726}, \href
  {https://ui.adsabs.harvard.edu/abs/2014MNRAS.445..225C} {445, 225}

\bibitem[\protect\citeauthoryear{{Corre} et~al.,}{{Corre}
  et~al.}{2018}]{Corre18}
{Corre} D.,  et~al., 2018, \mn@doi [\aap] {10.1051/0004-6361/201832926}, \href
  {https://ui.adsabs.harvard.edu/abs/2018A&A...617A.141C} {617, A141}

\bibitem[\protect\citeauthoryear{{D'Elia} et~al.,}{{D'Elia}
  et~al.}{2014}]{DElia14}
{D'Elia} V.,  et~al., 2014, \mn@doi [\aap] {10.1051/0004-6361/201323057}, \href
  {https://ui.adsabs.harvard.edu/abs/2014A&A...564A..38D} {564, A38}

\bibitem[\protect\citeauthoryear{{De Cia}, {Ledoux}, {Petitjean}  \&
  {Savaglio}}{{De Cia} et~al.}{2018}]{DeCia18}
{De Cia} A.,  {Ledoux} C.,  {Petitjean} P.,   {Savaglio} S.,  2018, \mn@doi
  [\aap] {10.1051/0004-6361/201731970}, \href
  {https://ui.adsabs.harvard.edu/abs/2018A&A...611A..76D} {611, A76}

\bibitem[\protect\citeauthoryear{{Decarli} et~al.,}{{Decarli}
  et~al.}{2016}]{Decarli16}
{Decarli} R.,  et~al., 2016, \mn@doi [\apj] {10.3847/1538-4357/833/1/70}, \href
  {https://ui.adsabs.harvard.edu/abs/2016ApJ...833...70D} {833, 70}

\bibitem[\protect\citeauthoryear{{Dessauges-Zavadsky}
  et~al.,}{{Dessauges-Zavadsky} et~al.}{2015}]{DessaugesZavadsky15}
{Dessauges-Zavadsky} M.,  et~al., 2015, \mn@doi [\aap]
  {10.1051/0004-6361/201424661}, \href
  {https://ui.adsabs.harvard.edu/abs/2015A&A...577A..50D} {577, A50}

\bibitem[\protect\citeauthoryear{{Endo} et~al.,}{{Endo} et~al.}{2007}]{Endo07}
{Endo} A.,  et~al., 2007, \mn@doi [\apj] {10.1086/512764}, \href
  {https://ui.adsabs.harvard.edu/abs/2007ApJ...659.1431E} {659, 1431}

\bibitem[\protect\citeauthoryear{{Francis} \& {Hewett}}{{Francis} \&
  {Hewett}}{1993}]{Francis93}
{Francis} P.~J.,  {Hewett} P.~C.,  1993, \mn@doi [\aj] {10.1086/116542}, \href
  {https://ui.adsabs.harvard.edu/abs/1993AJ....105.1633F} {105, 1633}

\bibitem[\protect\citeauthoryear{{Friis} et~al.,}{{Friis}
  et~al.}{2015}]{Friis15}
{Friis} M.,  et~al., 2015, \mn@doi [\mnras] {10.1093/mnras/stv960}, \href
  {https://ui.adsabs.harvard.edu/abs/2015MNRAS.451..167F} {451, 167}

\bibitem[\protect\citeauthoryear{{Fruchter} et~al.,}{{Fruchter}
  et~al.}{2006}]{Fruchter06}
{Fruchter} A.~S.,  et~al., 2006, \mn@doi [\nat] {10.1038/nature04787}, \href
  {https://ui.adsabs.harvard.edu/abs/2006Natur.441..463F} {441, 463}

\bibitem[\protect\citeauthoryear{{Fynbo}, {Ledoux}, {M{\o}ller}, {Thomsen}  \&
  {Burud}}{{Fynbo} et~al.}{2003}]{Fynbo03}
{Fynbo} J.~P.~U.,  {Ledoux} C.,  {M{\o}ller} P.,  {Thomsen} B.,   {Burud} I.,
  2003, \mn@doi [\aap] {10.1051/0004-6361:20030840}, \href
  {https://ui.adsabs.harvard.edu/abs/2003A&A...407..147F} {407, 147}

\bibitem[\protect\citeauthoryear{{Fynbo} et~al.,}{{Fynbo}
  et~al.}{2006}]{Fynbo06}
{Fynbo} J.~P.~U.,  et~al., 2006, \mn@doi [\aap] {10.1051/0004-6361:20065056},
  \href {https://ui.adsabs.harvard.edu/abs/2006A&A...451L..47F} {451, L47}

\bibitem[\protect\citeauthoryear{{Fynbo} et~al.,}{{Fynbo}
  et~al.}{2009}]{Fynbo09}
{Fynbo} J.~P.~U.,  et~al., 2009, \mn@doi [\apjs] {10.1088/0067-0049/185/2/526},
  \href {https://ui.adsabs.harvard.edu/abs/2009ApJS..185..526F} {185, 526}

\bibitem[\protect\citeauthoryear{{Fynbo} et~al.,}{{Fynbo}
  et~al.}{2018}]{Fynbo18}
{Fynbo} J.~P.~U.,  et~al., 2018, \mn@doi [\mnras] {10.1093/mnras/sty1520},
  \href {https://ui.adsabs.harvard.edu/abs/2018MNRAS.479.2126F} {479, 2126}

\bibitem[\protect\citeauthoryear{{Geach}, {Smail}, {Moran}, {MacArthur},
  {Lagos}  \& {Edge}}{{Geach} et~al.}{2011}]{Geach11}
{Geach} J.~E.,  {Smail} I.,  {Moran} S.~M.,  {MacArthur} L.~A.,  {Lagos} C.
  d.~P.,   {Edge} A.~C.,  2011, \mn@doi [\apjl] {10.1088/2041-8205/730/2/L19},
  \href {https://ui.adsabs.harvard.edu/abs/2011ApJ...730L..19G} {730, L19}

\bibitem[\protect\citeauthoryear{{Genzel} et~al.,}{{Genzel}
  et~al.}{2012}]{Genzel12}
{Genzel} R.,  et~al., 2012, \mn@doi [\apj] {10.1088/0004-637X/746/1/69}, \href
  {https://ui.adsabs.harvard.edu/abs/2012ApJ...746...69G} {746, 69}

\bibitem[\protect\citeauthoryear{{Genzel} et~al.,}{{Genzel}
  et~al.}{2015}]{Genzel15}
{Genzel} R.,  et~al., 2015, \mn@doi [\apj] {10.1088/0004-637X/800/1/20}, \href
  {https://ui.adsabs.harvard.edu/abs/2015ApJ...800...20G} {800, 20}

\bibitem[\protect\citeauthoryear{{Greiner} et~al.,}{{Greiner}
  et~al.}{2015}]{Greiner15}
{Greiner} J.,  et~al., 2015, \mn@doi [\apj] {10.1088/0004-637X/809/1/76}, \href
  {https://ui.adsabs.harvard.edu/abs/2015ApJ...809...76G} {809, 76}

\bibitem[\protect\citeauthoryear{{Hamanowicz} et~al.,}{{Hamanowicz}
  et~al.}{2020}]{Hamanowicz20}
{Hamanowicz} A.,  et~al., 2020, \mn@doi [\mnras] {10.1093/mnras/stz3590}, \href
  {https://ui.adsabs.harvard.edu/abs/2020MNRAS.492.2347H} {492, 2347}

\bibitem[\protect\citeauthoryear{{Hatsukade} et~al.,}{{Hatsukade}
  et~al.}{2007}]{Hatsukade07}
{Hatsukade} B.,  et~al., 2007, \mn@doi [\pasj] {10.1093/pasj/59.1.67}, \href
  {https://ui.adsabs.harvard.edu/abs/2007PASJ...59...67H} {59, 67}

\bibitem[\protect\citeauthoryear{{Hatsukade}, {Kohno}, {Endo}, {Nakanishi}  \&
  {Ohta}}{{Hatsukade} et~al.}{2011}]{Hatsukade11}
{Hatsukade} B.,  {Kohno} K.,  {Endo} A.,  {Nakanishi} K.,   {Ohta} K.,  2011,
  \mn@doi [\apj] {10.1088/0004-637X/738/1/33}, \href
  {https://ui.adsabs.harvard.edu/abs/2011ApJ...738...33H} {738, 33}

\bibitem[\protect\citeauthoryear{{Hatsukade}, {Ohta}, {Endo}, {Nakanishi},
  {Tamura}, {Hashimoto}  \& {Kohno}}{{Hatsukade} et~al.}{2014}]{Hatsukade14}
{Hatsukade} B.,  {Ohta} K.,  {Endo} A.,  {Nakanishi} K.,  {Tamura} Y.,
  {Hashimoto} T.,   {Kohno} K.,  2014, \mn@doi [\nat] {10.1038/nature13325},
  \href {https://ui.adsabs.harvard.edu/abs/2014Natur.510..247H} {510, 247}

\bibitem[\protect\citeauthoryear{{Hatsukade}, {Hashimoto}, {Kohno},
  {Nakanishi}, {Ohta}, {Niino}, {Tamura}  \& {T{\'o}th}}{{Hatsukade}
  et~al.}{2019}]{Hatsukade19}
{Hatsukade} B.,  {Hashimoto} T.,  {Kohno} K.,  {Nakanishi} K.,  {Ohta} K.,
  {Niino} Y.,  {Tamura} Y.,   {T{\'o}th} L.~V.,  2019, \mn@doi [\apj]
  {10.3847/1538-4357/ab1649}, \href
  {https://ui.adsabs.harvard.edu/abs/2019ApJ...876...91H} {876, 91}

\bibitem[\protect\citeauthoryear{{Hatsukade}, {Ohta}, {Hashimoto}, {Kohno},
  {Nakanishi}, {Niino}  \& {Tamura}}{{Hatsukade} et~al.}{2020}]{Hatsukade20}
{Hatsukade} B.,  {Ohta} K.,  {Hashimoto} T.,  {Kohno} K.,  {Nakanishi} K.,
  {Niino} Y.,   {Tamura} Y.,  2020, \mn@doi [\apj] {10.3847/1538-4357/ab7992},
  \href {https://ui.adsabs.harvard.edu/abs/2020ApJ...892...42H} {892, 42}

\bibitem[\protect\citeauthoryear{{Heintz} \& {Watson}}{{Heintz} \&
  {Watson}}{2020}]{HeintzWatson20}
{Heintz} K.~E.,  {Watson} D.,  2020, \mn@doi [\apjl]
  {10.3847/2041-8213/ab6733}, \href
  {https://ui.adsabs.harvard.edu/abs/2020ApJ...889L...7H} {889, L7}

\bibitem[\protect\citeauthoryear{{Heintz} et~al.,}{{Heintz}
  et~al.}{2019}]{Heintz19}
{Heintz} K.~E.,  et~al., 2019, \mn@doi [\aap] {10.1051/0004-6361/201936250},
  \href {https://ui.adsabs.harvard.edu/abs/2019A&A...629A.131H} {629, A131}

\bibitem[\protect\citeauthoryear{{Hjorth} et~al.,}{{Hjorth}
  et~al.}{2003}]{Hjorth03}
{Hjorth} J.,  et~al., 2003, \mn@doi [\nat] {10.1038/nature01750}, \href
  {https://ui.adsabs.harvard.edu/abs/2003Natur.423..847H} {423, 847}

\bibitem[\protect\citeauthoryear{{Inami} et~al.,}{{Inami}
  et~al.}{2020}]{Inami20}
{Inami} H.,  et~al., 2020, \mn@doi [\apj] {10.3847/1538-4357/abba2f}, \href
  {https://ui.adsabs.harvard.edu/abs/2020ApJ...902..113I} {902, 113}

\bibitem[\protect\citeauthoryear{{Israel}}{{Israel}}{1997}]{Israel97}
{Israel} F.~P.,  1997, \aap, \href
  {https://ui.adsabs.harvard.edu/abs/1997A&A...328..471I} {328, 471}

\bibitem[\protect\citeauthoryear{{Jakobsson} et~al.,}{{Jakobsson}
  et~al.}{2004}]{Jakobsson04}
{Jakobsson} P.,  et~al., 2004, \mn@doi [\aap] {10.1051/0004-6361:20041233},
  \href {https://ui.adsabs.harvard.edu/abs/2004A&A...427..785J} {427, 785}

\bibitem[\protect\citeauthoryear{{Jakobsson} et~al.,}{{Jakobsson}
  et~al.}{2006}]{Jakobsson06}
{Jakobsson} P.,  et~al., 2006, \mn@doi [\aap] {10.1051/0004-6361:20066405},
  \href {https://ui.adsabs.harvard.edu/abs/2006A&A...460L..13J} {460, L13}

\bibitem[\protect\citeauthoryear{{Kacprzak}, {Murphy}  \&
  {Churchill}}{{Kacprzak} et~al.}{2010}]{Kacprzak10}
{Kacprzak} G.~G.,  {Murphy} M.~T.,   {Churchill} C.~W.,  2010, \mn@doi [\mnras]
  {10.1111/j.1365-2966.2010.16667.x}, \href
  {https://ui.adsabs.harvard.edu/abs/2010MNRAS.406..445K} {406, 445}

\bibitem[\protect\citeauthoryear{{Kanekar}, {Prochaska}, {Neeleman},
  {Christensen}, {M{\o}ller}, {Zwaan}, {Fynbo}  \&
  {Dessauges-Zavadsky}}{{Kanekar} et~al.}{2020}]{Kanekar20}
{Kanekar} N.,  {Prochaska} J.~X.,  {Neeleman} M.,  {Christensen} L.,
  {M{\o}ller} P.,  {Zwaan} M.~A.,  {Fynbo} J.~P.~U.,   {Dessauges-Zavadsky} M.,
   2020, \mn@doi [\apjl] {10.3847/2041-8213/abb4e1}, \href
  {https://ui.adsabs.harvard.edu/abs/2020ApJ...901L...5K} {901, L5}

\bibitem[\protect\citeauthoryear{{Kistler}, {Y{\"u}ksel}, {Beacom}, {Hopkins}
  \& {Wyithe}}{{Kistler} et~al.}{2009}]{Kistler09}
{Kistler} M.~D.,  {Y{\"u}ksel} H.,  {Beacom} J.~F.,  {Hopkins} A.~M.,
  {Wyithe} J. S.~B.,  2009, \mn@doi [\apjl] {10.1088/0004-637X/705/2/L104},
  \href {https://ui.adsabs.harvard.edu/abs/2009ApJ...705L.104K} {705, L104}

\bibitem[\protect\citeauthoryear{{Klitsch}, {P{\'e}roux}, {Zwaan}, {Smail},
  {Oteo}, {Biggs}, {Popping}  \& {Swinbank}}{{Klitsch}
  et~al.}{2018}]{Klitsch18}
{Klitsch} A.,  {P{\'e}roux} C.,  {Zwaan} M.~A.,  {Smail} I.,  {Oteo} I.,
  {Biggs} A.~D.,  {Popping} G.,   {Swinbank} A.~M.,  2018, \mn@doi [\mnras]
  {10.1093/mnras/stx3184}, \href
  {https://ui.adsabs.harvard.edu/abs/2018MNRAS.475..492K} {475, 492}

\bibitem[\protect\citeauthoryear{{Kohno} et~al.,}{{Kohno}
  et~al.}{2005}]{Kohno05}
{Kohno} K.,  et~al., 2005, \mn@doi [\pasj] {10.1093/pasj/57.1.147}, \href
  {https://ui.adsabs.harvard.edu/abs/2005PASJ...57..147K} {57, 147}

\bibitem[\protect\citeauthoryear{{Kr{\"u}hler} et~al.,}{{Kr{\"u}hler}
  et~al.}{2013}]{Kruhler13}
{Kr{\"u}hler} T.,  et~al., 2013, \mn@doi [\aap] {10.1051/0004-6361/201321772},
  \href {https://ui.adsabs.harvard.edu/abs/2013A&A...557A..18K} {557, A18}

\bibitem[\protect\citeauthoryear{{Kr{\"u}hler} et~al.,}{{Kr{\"u}hler}
  et~al.}{2015}]{Kruhler15}
{Kr{\"u}hler} T.,  et~al., 2015, \mn@doi [\aap] {10.1051/0004-6361/201425561},
  \href {https://ui.adsabs.harvard.edu/abs/2015A&A...581A.125K} {581, A125}

\bibitem[\protect\citeauthoryear{{Ledoux}, {Vreeswijk}, {Smette}, {Fox},
  {Petitjean}, {Ellison}, {Fynbo}  \& {Savaglio}}{{Ledoux}
  et~al.}{2009}]{Ledoux09}
{Ledoux} C.,  {Vreeswijk} P.~M.,  {Smette} A.,  {Fox} A.~J.,  {Petitjean} P.,
  {Ellison} S.~L.,  {Fynbo} J.~P.~U.,   {Savaglio} S.,  2009, \mn@doi [\aap]
  {10.1051/0004-6361/200811572}, \href
  {https://ui.adsabs.harvard.edu/abs/2009A&A...506..661L} {506, 661}

\bibitem[\protect\citeauthoryear{{Leroy} et~al.,}{{Leroy}
  et~al.}{2011}]{Leroy11}
{Leroy} A.~K.,  et~al., 2011, \mn@doi [\apj] {10.1088/0004-637X/737/1/12},
  \href {https://ui.adsabs.harvard.edu/abs/2011ApJ...737...12L} {737, 12}

\bibitem[\protect\citeauthoryear{{Maiolino} et~al.,}{{Maiolino}
  et~al.}{2008}]{Maiolino08}
{Maiolino} R.,  et~al., 2008, \mn@doi [\aap] {10.1051/0004-6361:200809678},
  \href {https://ui.adsabs.harvard.edu/abs/2008A&A...488..463M} {488, 463}

\bibitem[\protect\citeauthoryear{{McMullin}, {Waters}, {Schiebel}, {Young}  \&
  {Golap}}{{McMullin} et~al.}{2007}]{McMullin07}
{McMullin} J.~P.,  {Waters} B.,  {Schiebel} D.,  {Young} W.,   {Golap} K.,
  2007, in {Shaw} R.~A.,  {Hill} F.,   {Bell} D.~J.,  eds,  Astronomical
  Society of the Pacific Conference Series Vol. 376, Astronomical Data Analysis
  Software and Systems XVI. p.~127

\bibitem[\protect\citeauthoryear{{Micha{\l}owski} et~al.,}{{Micha{\l}owski}
  et~al.}{2016}]{Michalowski16}
{Micha{\l}owski} M.~J.,  et~al., 2016, \mn@doi [\aap]
  {10.1051/0004-6361/201629441}, \href
  {https://ui.adsabs.harvard.edu/abs/2016A&A...595A..72M} {595, A72}

\bibitem[\protect\citeauthoryear{{Micha{\l}owski} et~al.,}{{Micha{\l}owski}
  et~al.}{2018}]{Michalowski18}
{Micha{\l}owski} M.~J.,  et~al., 2018, \mn@doi [\aap]
  {10.1051/0004-6361/201833250}, \href
  {https://ui.adsabs.harvard.edu/abs/2018A&A...617A.143M} {617, A143}

\bibitem[\protect\citeauthoryear{{M{\o}ller} \& {Warren}}{{M{\o}ller} \&
  {Warren}}{1993}]{Moller93}
{M{\o}ller} P.,  {Warren} S.~J.,  1993, \aap, \href
  {https://ui.adsabs.harvard.edu/abs/1993A&A...270...43M} {270, 43}

\bibitem[\protect\citeauthoryear{{M{\o}ller}, {Fynbo}, {Ledoux}  \&
  {Nilsson}}{{M{\o}ller} et~al.}{2013}]{Moller13}
{M{\o}ller} P.,  {Fynbo} J.~P.~U.,  {Ledoux} C.,   {Nilsson} K.~K.,  2013,
  \mn@doi [\mnras] {10.1093/mnras/stt067}, \href
  {https://ui.adsabs.harvard.edu/abs/2013MNRAS.430.2680M} {430, 2680}

\bibitem[\protect\citeauthoryear{{Neeleman}, {Wolfe}, {Prochaska}  \&
  {Rafelski}}{{Neeleman} et~al.}{2013}]{Neeleman13}
{Neeleman} M.,  {Wolfe} A.~M.,  {Prochaska} J.~X.,   {Rafelski} M.,  2013,
  \mn@doi [\apj] {10.1088/0004-637X/769/1/54}, \href
  {https://ui.adsabs.harvard.edu/abs/2013ApJ...769...54N} {769, 54}

\bibitem[\protect\citeauthoryear{{Neeleman}, {Kanekar}, {Prochaska},
  {Christensen}, {Dessauges-Zavadsky}, {Fynbo}, {M{\o}ller}  \&
  {Zwaan}}{{Neeleman} et~al.}{2018}]{Neeleman18}
{Neeleman} M.,  {Kanekar} N.,  {Prochaska} J.~X.,  {Christensen} L.,
  {Dessauges-Zavadsky} M.,  {Fynbo} J. P.~U.,  {M{\o}ller} P.,   {Zwaan} M.~A.,
   2018, \mn@doi [\apjl] {10.3847/2041-8213/aab5b1}, \href
  {https://ui.adsabs.harvard.edu/abs/2018ApJ...856L..12N} {856, L12}

\bibitem[\protect\citeauthoryear{{Noterdaeme}, {Petitjean}, {Ledoux}  \&
  {Srianand}}{{Noterdaeme} et~al.}{2009}]{Noterdaeme09}
{Noterdaeme} P.,  {Petitjean} P.,  {Ledoux} C.,   {Srianand} R.,  2009, \mn@doi
  [\aap] {10.1051/0004-6361/200912768}, \href
  {https://ui.adsabs.harvard.edu/abs/2009A&A...505.1087N} {505, 1087}

\bibitem[\protect\citeauthoryear{{Pavesi} et~al.,}{{Pavesi}
  et~al.}{2018}]{Pavesi18}
{Pavesi} R.,  et~al., 2018, \mn@doi [\apj] {10.3847/1538-4357/aacb79}, \href
  {https://ui.adsabs.harvard.edu/abs/2018ApJ...864...49P} {864, 49}

\bibitem[\protect\citeauthoryear{{Prochaska}, {Chen}, {Dessauges-Zavadsky}  \&
  {Bloom}}{{Prochaska} et~al.}{2007}]{Prochaska07}
{Prochaska} J.~X.,  {Chen} H.-W.,  {Dessauges-Zavadsky} M.,   {Bloom} J.~S.,
  2007, \mn@doi [\apj] {10.1086/520042}, \href
  {https://ui.adsabs.harvard.edu/abs/2007ApJ...666..267P} {666, 267}

\bibitem[\protect\citeauthoryear{{Prochaska} et~al.,}{{Prochaska}
  et~al.}{2009}]{Prochaska09}
{Prochaska} J.~X.,  et~al., 2009, \mn@doi [\apjl]
  {10.1088/0004-637X/691/1/L27}, \href
  {https://ui.adsabs.harvard.edu/abs/2009ApJ...691L..27P} {691, L27}

\bibitem[\protect\citeauthoryear{{Rafelski}, {Wolfe}, {Prochaska}, {Neeleman}
  \& {Mendez}}{{Rafelski} et~al.}{2012}]{Rafelski12}
{Rafelski} M.,  {Wolfe} A.~M.,  {Prochaska} J.~X.,  {Neeleman} M.,   {Mendez}
  A.~J.,  2012, \mn@doi [\apj] {10.1088/0004-637X/755/2/89}, \href
  {https://ui.adsabs.harvard.edu/abs/2012ApJ...755...89R} {755, 89}

\bibitem[\protect\citeauthoryear{{Rahmani} et~al.,}{{Rahmani}
  et~al.}{2018}]{Rahmani18}
{Rahmani} H.,  et~al., 2018, \mn@doi [\mnras] {10.1093/mnras/stx2726}, \href
  {https://ui.adsabs.harvard.edu/abs/2018MNRAS.474..254R} {474, 254}

\bibitem[\protect\citeauthoryear{{Ranjan}, {Noterdaeme}, {Krogager},
  {Petitjean}, {Srianand}, {Balashev}, {Gupta}  \& {Ledoux}}{{Ranjan}
  et~al.}{2020}]{Ranjan20}
{Ranjan} A.,  {Noterdaeme} P.,  {Krogager} J.~K.,  {Petitjean} P.,  {Srianand}
  R.,  {Balashev} S.~A.,  {Gupta} N.,   {Ledoux} C.,  2020, \mn@doi [\aap]
  {10.1051/0004-6361/201936078}, \href
  {https://ui.adsabs.harvard.edu/abs/2020A&A...633A.125R} {633, A125}

\bibitem[\protect\citeauthoryear{{Riechers} et~al.,}{{Riechers}
  et~al.}{2019}]{Riechers19}
{Riechers} D.~A.,  et~al., 2019, \mn@doi [\apj] {10.3847/1538-4357/aafc27},
  \href {https://ui.adsabs.harvard.edu/abs/2019ApJ...872....7R} {872, 7}

\bibitem[\protect\citeauthoryear{{Robertson} \& {Ellis}}{{Robertson} \&
  {Ellis}}{2012}]{Robertson12}
{Robertson} B.~E.,  {Ellis} R.~S.,  2012, \mn@doi [\apj]
  {10.1088/0004-637X/744/2/95}, \href
  {https://ui.adsabs.harvard.edu/abs/2012ApJ...744...95R} {744, 95}

\bibitem[\protect\citeauthoryear{{Savaglio} et~al.,}{{Savaglio}
  et~al.}{2005}]{Savaglio05}
{Savaglio} S.,  et~al., 2005, \mn@doi [\apj] {10.1086/497331}, \href
  {https://ui.adsabs.harvard.edu/abs/2005ApJ...635..260S} {635, 260}

\bibitem[\protect\citeauthoryear{{Selsing} et~al.,}{{Selsing}
  et~al.}{2019}]{Selsing19}
{Selsing} J.,  et~al., 2019, \mn@doi [\aap] {10.1051/0004-6361/201832835},
  \href {https://ui.adsabs.harvard.edu/abs/2019A&A...623A..92S} {623, A92}

\bibitem[\protect\citeauthoryear{{Solomon} \& {Vanden Bout}}{{Solomon} \&
  {Vanden Bout}}{2005}]{Solomon05}
{Solomon} P.~M.,  {Vanden Bout} P.~A.,  2005, \mn@doi [\araa]
  {10.1146/annurev.astro.43.051804.102221}, \href
  {https://ui.adsabs.harvard.edu/abs/2005ARA&A..43..677S} {43, 677}

\bibitem[\protect\citeauthoryear{{Stanek} et~al.,}{{Stanek}
  et~al.}{2003}]{Stanek03}
{Stanek} K.~Z.,  et~al., 2003, \mn@doi [\apjl] {10.1086/376976}, \href
  {https://ui.adsabs.harvard.edu/abs/2003ApJ...591L..17S} {591, L17}

\bibitem[\protect\citeauthoryear{{Stanway}, {Bremer}, {Tanvir}, {Levan}  \&
  {Davies}}{{Stanway} et~al.}{2011}]{Stanway11}
{Stanway} E.~R.,  {Bremer} M.~N.,  {Tanvir} N.~R.,  {Levan} A.~J.,   {Davies}
  L. J.~M.,  2011, \mn@doi [\mnras] {10.1111/j.1365-2966.2010.17534.x}, \href
  {https://ui.adsabs.harvard.edu/abs/2011MNRAS.410.1496S} {410, 1496}

\bibitem[\protect\citeauthoryear{{Stanway}, {Levan}, {Tanvir}, {Wiersema}  \&
  {van der Laan}}{{Stanway} et~al.}{2015}]{Stanway15}
{Stanway} E.~R.,  {Levan} A.~J.,  {Tanvir} N.~R.,  {Wiersema} K.,   {van der
  Laan} T.~P.~R.,  2015, \mn@doi [\apjl] {10.1088/2041-8205/798/1/L7}, \href
  {https://ui.adsabs.harvard.edu/abs/2015ApJ...798L...7S} {798, L7}

\bibitem[\protect\citeauthoryear{{Svensson}, {Levan}, {Tanvir}, {Fruchter}  \&
  {Strolger}}{{Svensson} et~al.}{2010}]{Svensson10}
{Svensson} K.~M.,  {Levan} A.~J.,  {Tanvir} N.~R.,  {Fruchter} A.~S.,
  {Strolger} L.~G.,  2010, \mn@doi [\mnras] {10.1111/j.1365-2966.2010.16442.x},
  \href {https://ui.adsabs.harvard.edu/abs/2010MNRAS.405...57S} {405, 57}

\bibitem[\protect\citeauthoryear{{Tacconi} et~al.,}{{Tacconi}
  et~al.}{2010}]{Tacconi10}
{Tacconi} L.~J.,  et~al., 2010, \mn@doi [\nat] {10.1038/nature08773}, \href
  {https://ui.adsabs.harvard.edu/abs/2010Natur.463..781T} {463, 781}

\bibitem[\protect\citeauthoryear{{Tacconi} et~al.,}{{Tacconi}
  et~al.}{2013}]{Tacconi13}
{Tacconi} L.~J.,  et~al., 2013, \mn@doi [\apj] {10.1088/0004-637X/768/1/74},
  \href {https://ui.adsabs.harvard.edu/abs/2013ApJ...768...74T} {768, 74}

\bibitem[\protect\citeauthoryear{{Tacconi} et~al.,}{{Tacconi}
  et~al.}{2018}]{Tacconi18}
{Tacconi} L.~J.,  et~al., 2018, \mn@doi [\apj] {10.3847/1538-4357/aaa4b4},
  \href {https://ui.adsabs.harvard.edu/abs/2018ApJ...853..179T} {853, 179}

\bibitem[\protect\citeauthoryear{{Tiley}, {Bureau}, {Saintonge}, {Topal},
  {Davis}  \& {Torii}}{{Tiley} et~al.}{2016}]{Tiley16}
{Tiley} A.~L.,  {Bureau} M.,  {Saintonge} A.,  {Topal} S.,  {Davis} T.~A.,
  {Torii} K.,  2016, \mn@doi [\mnras] {10.1093/mnras/stw1545}, \href
  {https://ui.adsabs.harvard.edu/abs/2016MNRAS.461.3494T} {461, 3494}

\bibitem[\protect\citeauthoryear{{Tumlinson}, {Prochaska}, {Chen},
  {Dessauges-Zavadsky}  \& {Bloom}}{{Tumlinson} et~al.}{2007}]{Tumlinson07}
{Tumlinson} J.,  {Prochaska} J.~X.,  {Chen} H.-W.,  {Dessauges-Zavadsky} M.,
  {Bloom} J.~S.,  2007, \mn@doi [\apj] {10.1086/521294}, \href
  {https://ui.adsabs.harvard.edu/abs/2007ApJ...668..667T} {668, 667}

\bibitem[\protect\citeauthoryear{{Valentino} et~al.,}{{Valentino}
  et~al.}{2018}]{Valentino18}
{Valentino} F.,  et~al., 2018, \mn@doi [\apj] {10.3847/1538-4357/aaeb88}, \href
  {https://ui.adsabs.harvard.edu/abs/2018ApJ...869...27V} {869, 27}

\bibitem[\protect\citeauthoryear{{Valentino} et~al.,}{{Valentino}
  et~al.}{2020}]{Valentino20}
{Valentino} F.,  et~al., 2020, \mn@doi [\apj] {10.3847/1538-4357/ab6603}, \href
  {https://ui.adsabs.harvard.edu/abs/2020ApJ...890...24V} {890, 24}

\bibitem[\protect\citeauthoryear{{Vreeswijk} et~al.,}{{Vreeswijk}
  et~al.}{2004}]{Vreeswijk04}
{Vreeswijk} P.~M.,  et~al., 2004, \mn@doi [\aap] {10.1051/0004-6361:20040086},
  \href {https://ui.adsabs.harvard.edu/abs/2004A&A...419..927V} {419, 927}

\bibitem[\protect\citeauthoryear{{Vreeswijk} et~al.,}{{Vreeswijk}
  et~al.}{2007}]{Vreeswijk07}
{Vreeswijk} P.~M.,  et~al., 2007, \mn@doi [\aap] {10.1051/0004-6361:20066780},
  \href {https://ui.adsabs.harvard.edu/abs/2007A&A...468...83V} {468, 83}

\bibitem[\protect\citeauthoryear{{Walter}, {Wei{\ss}}, {Downes}, {Decarli}  \&
  {Henkel}}{{Walter} et~al.}{2011}]{Walter11}
{Walter} F.,  {Wei{\ss}} A.,  {Downes} D.,  {Decarli} R.,   {Henkel} C.,  2011,
  \mn@doi [\apj] {10.1088/0004-637X/730/1/18}, \href
  {https://ui.adsabs.harvard.edu/abs/2011ApJ...730...18W} {730, 18}

\bibitem[\protect\citeauthoryear{{Wang}, {Chen}  \& {Huang}}{{Wang}
  et~al.}{2012}]{Wang12}
{Wang} W.-H.,  {Chen} H.-W.,   {Huang} K.-Y.,  2012, \mn@doi [\apjl]
  {10.1088/2041-8205/761/2/L32}, \href
  {https://ui.adsabs.harvard.edu/abs/2012ApJ...761L..32W} {761, L32}

\bibitem[\protect\citeauthoryear{{Wijers}, {Bloom}, {Bagla}  \&
  {Natarajan}}{{Wijers} et~al.}{1998}]{Wijers98}
{Wijers} R. A.~M.~J.,  {Bloom} J.~S.,  {Bagla} J.~S.,   {Natarajan} P.,  1998,
  \mn@doi [\mnras] {10.1046/j.1365-8711.1998.01328.x}, \href
  {https://ui.adsabs.harvard.edu/abs/1998MNRAS.294L..13W} {294, L13}

\bibitem[\protect\citeauthoryear{{Woosley} \& {Bloom}}{{Woosley} \&
  {Bloom}}{2006}]{Woosley06}
{Woosley} S.~E.,  {Bloom} J.~S.,  2006, \mn@doi [\araa]
  {10.1146/annurev.astro.43.072103.150558}, \href
  {https://ui.adsabs.harvard.edu/abs/2006ARA&A..44..507W} {44, 507}

\bibitem[\protect\citeauthoryear{{de Ugarte Postigo} et~al.,}{{de Ugarte
  Postigo} et~al.}{2020}]{deUgartePostigo20}
{de Ugarte Postigo} A.,  et~al., 2020, \mn@doi [\aap]
  {10.1051/0004-6361/201936668}, \href
  {https://ui.adsabs.harvard.edu/abs/2020A&A...633A..68D} {633, A68}

\makeatother
\end{thebibliography}

\bsp	
\label{lastpage}
\end{document}